\documentclass{elsart4-1}

\usepackage{amssymb}
\usepackage{amsfonts}
\usepackage{graphicx}
\usepackage[english,francais]{babel}

\begin{document}
\centerline{Physics or Astrophysics/Header}
\begin{frontmatter}
\selectlanguage{english}

\title{Experimental study of work fluctuations in a harmonic oscillator}
\author{S. Joubaud},
\author{N. B. Garnier},
\author{F. Douarche},
\author{A. Petrosyan},
\author{S. Ciliberto}
\ead{sergio.ciliberto@ens-lyon.fr}
\address{Laboratoire de Physique de l'ENS Lyon, CNRS UMR 5672,\\
        46, All\'ee d'Italie, 69364 Lyon CEDEX 07, France}


%



\begin{abstract}
The work fluctuations of a harmonic  oscillator in contact with a
thermostat and driven out of equilibrium by an external force are
studied experimentally. For the work both the transient and
stationary state fluctuation theorems hold. The finite time
corrections are very different from those of a first order
Langevin equation. The heat and work fluctuations are studied when
a periodic forcing is applied to the oscillator. The importance of
the choice of the ''good work'' to compute the free energy from
the Jarzinsky equality is discussed.

\vskip 0.5\baselineskip

\selectlanguage{francais} \noindent{\bf R\'esum\'e} \vskip
0.5\baselineskip \noindent {\bf Etude exp\'erimentale  des
fluctuations du travail dans un oscillateur harmonique. } Nous
\'etudions exp\'erimentalement les fluctuations du travail d'une
force externe sur un oscillateur harmonique en contact avec un
bain thermique. Nous trouvons que les th\'eor\`emes de
fluctuations pour les \'etats transitoires et stationnaires sont
v\'erifi\'es. Toutefois les corrections de temps fini pour
l'\'etat stationnaire sont tr\`es diff\'erentes de celles
calcul\'ees dans le cadre de l'\'equation de Langevin du premier
ordre. Nous \'etudions aussi le for\c{c}age p\'eriodique de
l'oscillateur. Enfin nous discutons l'importance du choix du ''bon
travail'' pour calculer l'\'energie libre en partant de
l'\'egalit\'e de Jarzinsky.

 \vskip
0.5\baselineskip \noindent{\small{\it Mots-cl\'es~:} th\'eor\`emes
de fluctuation~; travail; hors-\'equilibre}}

\end{abstract}

\end{frontmatter}

\selectlanguage{english}

\section{Introduction}
Thermal fluctuations play a very important role in the out of
equilibrium dynamics of small systems such those that one can find
in nanotechnologies and biophysics. Indeed in these systems the
amount of injected energy   is often comparable to that of thermal
fluctuations, which may produce unwanted and unexpected behaviors.
For example  very large fluctuations may force the system to move
in the opposite  direction  of the one  imposed by the external
forces. In the same way the heat exchanges with the thermal bath
are such that the heat may instantaneously flow from a cold source
to a hot one. Of course these are not  violations of the second
principle of  thermodynamics which fixes only the mean values and
it does not make any statement for fluctuations. The recent
fluctuation theorems FTs
\cite{Evansetal93}-\cite{Cohen}
 allow one to compute the probability of these negative
 events for the work and for the heat flux. It is interesting to
notice that the fluctuations of the work done by the  external
forces to drive the system between two equilibrium states  A and B
allows one to compute, in some cases,  the free energy difference
$\Delta F$ between $A$ and $B$, using either the Jarzinsky
equality (JE) \cite{jarzynski1,jarzynski2} or the Crooks relation
(CR)\cite{crooks1}. Indeed the JE and CR take advantage of these
work fluctuations and relate $\Delta F$ to the probability
distribution function (PDF) of the work performed on the system to
drive it from $A$ to $B$ along any path $\gamma$(either reversible
or irreversible) in the system parameter space. In this paper we
will study the possibility of using these theorems on real
experimental data. Indeed  the proofs of FTs and JE  are based on
a certain number of hypothesis; experimenting on real devices is
useful not only to check those hypothesis, but also to check
whether the predicted effects are observable or remain only a
theoretical tool. There are not many experimental tests of FTs.
Some of them are performed in dynamical systems
\cite{otherexperiment} in which the interpretation of the results
is very difficult. Other experiments are performed on stochastic
systems, one on a Brownian particle in a moving optical
trap~\cite{Wangetal:02:05} and another on electrical circuits
driven out of equilibrium by injecting in it a small
current~\cite{Garnier}. The last two systems are described by
first order Langevin equations and the results agree with the
predictions of ref.\cite{Cohen,Cohen1}.In this paper we study
 the out-of-equilibrium fluctuations of a
harmonic oscillator  whose damping is mainly produced by the
viscosity of the surrounding fluid, which acts as a thermal bath
of temperature $T$. The test using a harmonic oscillator is
particularly important because the harmonic oscillator is the
basis of many physical processes \cite{Zamponi:05,Douarche2006}.
The paper is organized as follow. In the next section we will
briefly describe the experimental set-up. The fluctuation theorem
will be studied is section three. In section 4 we will describe
the results for the work and heat for a the stationary state
fluctuation theorem when the oscillator is submitted to a periodic
driving. In section 5 the JE and Crooks are used to compute the
free energy. The importance of the choice of the ''good work'' for
the JE is pointed out too. We show indeed that a definition of
work which satisfies FT cannot be used to compute the free energy
with JE. Finally we conclude in section 6.

\begin{figure}
\centerline{\includegraphics[width=1.0\linewidth]{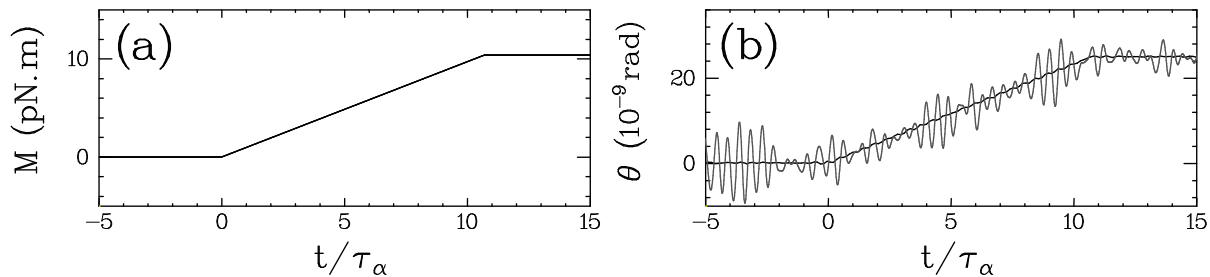}}
\caption{ a) Typical driving torque applied to the oscillator; b)
 Response of the oscillator to the external torque (gray line).
 The dark line represents the mean response $\bar \theta(t)$
 to the applied torque $M(t)$.}
\label{driver_fluct}
\end{figure}

\section{Experimental set-up}
We recall here only the main features of the experimental set-up,
more details can be found in ref.\cite{rsi,DouarcheJSM}. The
oscillator is a torsion pendulum composed by a brass wire and a
glass mirror glued in the middle of this wire. It is enclosed in a
cell filled by a water-glycerol solution at $60\%$ concentration.
The motion of this pendulum can be described by a second order
Langevin equation:
\begin{equation}
I_{\mathrm{eff}}\,\frac{{\rm d}^2{\theta}}{{\rm d}t^2} + \nu
\,\frac{{\rm d}{\theta}}{{\rm d}t}  + C\,\theta = M + \sqrt{2k_BT
\ \nu}\ \  \eta, \label{eqoscillator}
\end{equation}
where $\theta$ is the angular displacement of the pendulum,
$I_{\mathrm{eff}}$ is the total moment of inertia of the displaced
masses, $\nu$ is the oscillator viscous damping, $C$ is the
elastic torsional stiffness of the wire, $M$ is the external
torque, $k_B$ the Boltzmann constant and $\eta$ the noise,
delta-correlated in time. In our system the measured parameters
are the stiffness $C = 4.5 \times 10^{-4}
\textrm{N\,m\,rad}^{-1}$, the resonant frequency
$f_o=\sqrt{C/I_{\rm eff}}/(2\pi)=217$Hz and the relaxation time
$\tau_\alpha=2I_{\rm eff}/\nu= 9.5$ms. The external torque $M$ is
applied by means of a tiny electric current $J$ flowing in a coil
glued behind the mirror. The coil is inside a static magnetic
field, hence $M\propto J$. The measurement of $\theta$ is
performed by a differential interferometer, which uses two  laser
beams impinging on the pendulum mirror~\cite{rsi,DouarcheJSM}. The
measurement noise is two orders of magnitude smaller than the
thermal fluctuations of the pendulum. $\theta(t)$ is acquired with
a resolution of 24 bits at a sampling rate of $8192$Hz, which is
about 40 times $f_o$. The calibration accuracy of the apparatus,
tested at $M=0$ using the the Fluctuation Dissipation Theorem, is
better than $3\%$(see~\cite{DouarcheJSM}).
\begin{figure}
\centerline{\includegraphics[width=1.0\linewidth]{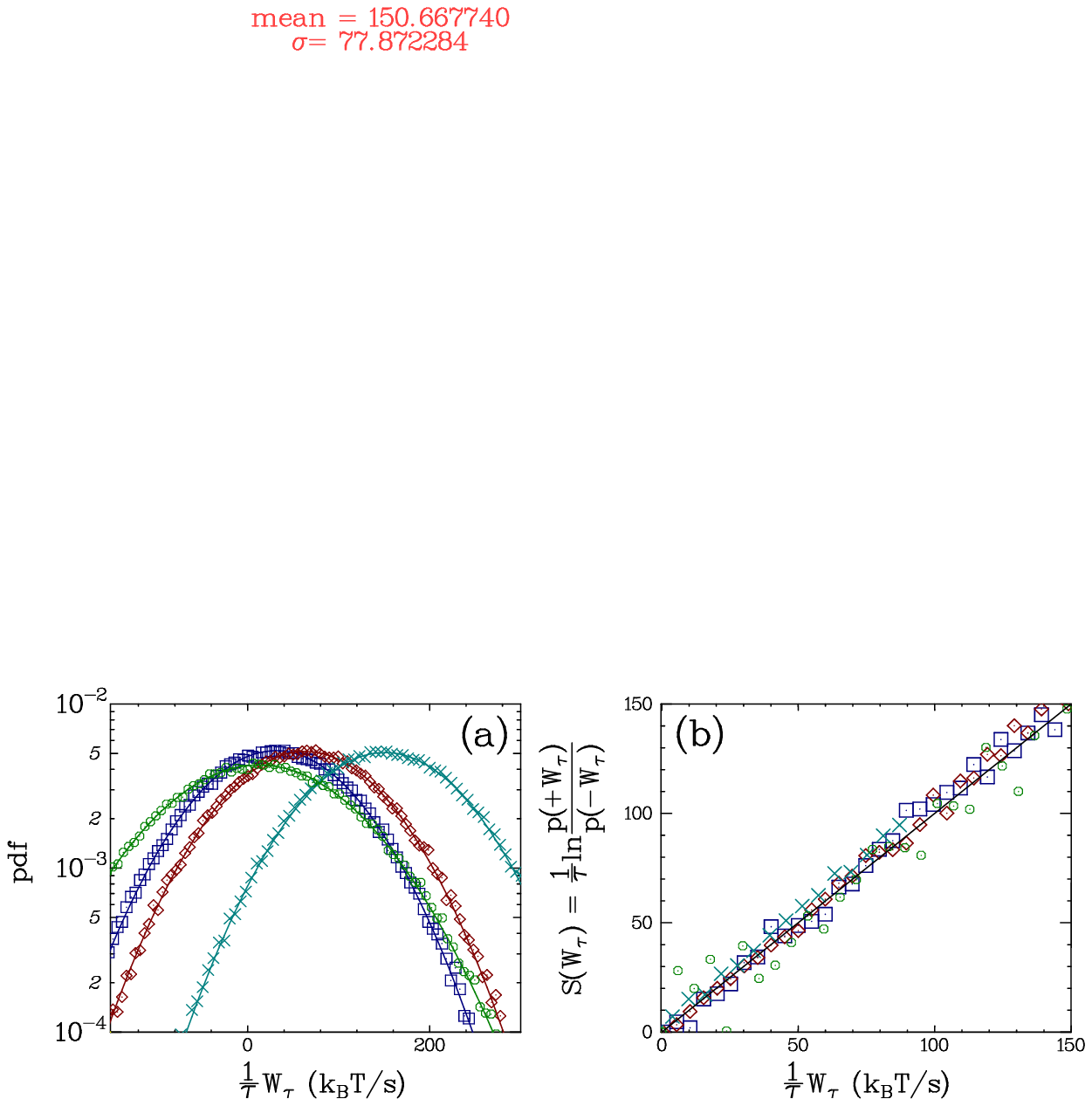}}
\caption{TFT.
   a) $P(W_\tau)$ for TFT for various $\tau/\tau_\alpha$: 0.31 $(\circ)$,
   1.015 $(\Box)$, 2.09 $(\diamond)$ and 4.97 $(\times)$. Continuous lines
   are Gaussian fits.
   b) TFT; $S(W_\tau)$ computed with the PDF of a).
   The straight continuous line has slope 1,
    {\em i.e.}, $\Sigma(\tau)=1, \ \forall \tau$.
   }
\label{fig:TFT}
\end{figure}

\section{Fluctuation theorems}
\label{section_ramp}

It is important to make a distinction between {\em Stationary
State Fluctuation Theorem} (SSFT) and and {\em Transient
Fluctuation Theorem} (TFT) (as defined in
ref.\cite{Cohen,Cohen1}). TFT implies that the initial state when
the force is applied to the system is an equilibrium one. In
contrast SSFT concerns out of equilibrium stationary states where
all transients have already relaxed.
 To study  SSFT and  TFT
we apply to the oscillator a time dependent torque $M(t)$ as
depicted in Fig.~\ref{driver_fluct}a, and we consider the work
$W_\tau$ done by $M(t)$ over a time $\tau$:
\begin{equation}
W_\tau= {1 \over k_B \ T} \ \int_{t_i}^{t_i+\tau} \left[ M(t)-M(t_i) \right]
{d \theta \over dt} dt \,.
\label{work}
\end{equation}
By definition, the TFT implies  that  $t_i=0$ in eq.\ref{work}
whereas $t_i \ge 3\tau_\alpha$ for SSFT. As a second choice for
$M(t)$, the linear ramp with a rising time $\tau_r$ is replaced by
a sinusoidal forcing; this leads to a new form of stationary state
which has never been considered in the context of FT. We examine
first the linear forcing $M(t)={M_o t / \tau_r}$
(Fig.\,\ref{driver_fluct}a)), with $M_o=10.4$ pN.m and $\tau_r$ =
0.1 s = 10.7 $\tau_\alpha$. The response of the oscillator to this
excitation is comparable to the thermal noise amplitude, as can be
seen in Fig.\,\ref{driver_fluct}b) where $\theta(t)$ is plotted
during the same time interval of Fig.\,\ref{driver_fluct}a).
Because of thermal noise the power $W_\tau$ injected into the
system (eq.\ref{work}) is itself a strongly fluctuating quantity.

The FTs state \cite{Cohen} that the probability density functions
(PDF) $P(W_\tau)$ of $W_\tau$ satisfies:
\begin{equation}
 \ln\left[{P(W_\tau)\over P(-W_\tau)}\right]= \Sigma(\tau)~W_\tau
\label{eq_FT1}
\end{equation}
where for  TFT
\begin{equation}
\Sigma(\tau)=1, ~  \forall \tau \label{eq_TFT}
\end{equation}
and for  SSFT
\begin{equation}
 \Sigma(\tau)\rightarrow \ 1 \ \ \ {\rm for} \ \ \
\tau\rightarrow \infty, \label{eq_SSFT}
\end{equation}
(see refs.\cite{Cohen1,Cohen} for a clear discussion on the
difference between TFT and SSFT).

\begin{figure}[!h]
\centerline{\includegraphics[width=1.0\linewidth]{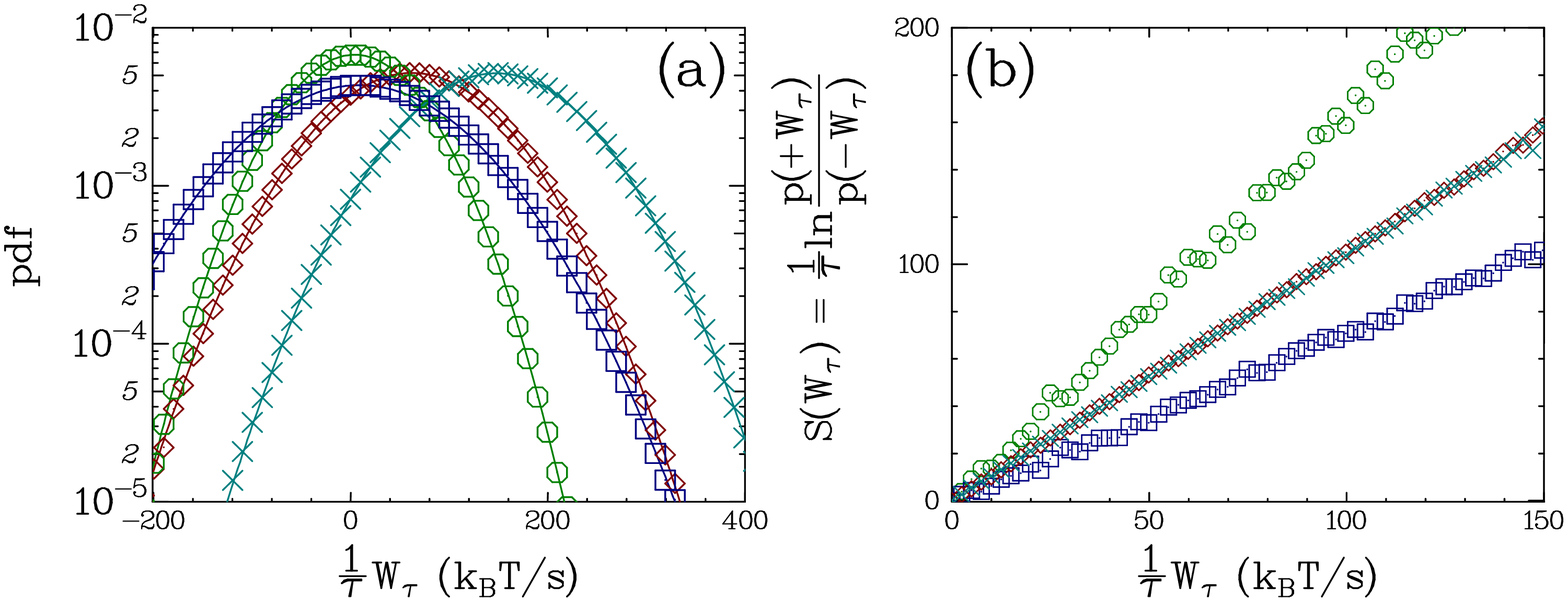}}
\centerline{\includegraphics[width=0.8\linewidth]{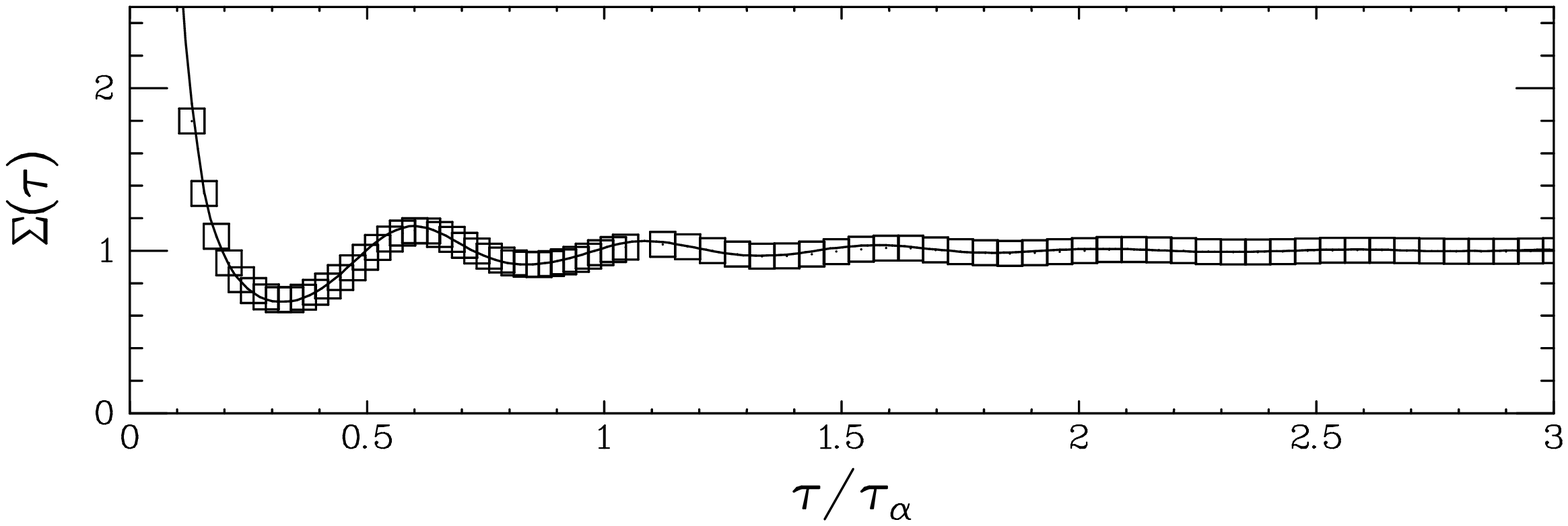}}
\caption{SSFT with a ramp forcing. a) PDF of $W_\tau$ for various
$\tau/\tau_\alpha$: 0.019 $(\circ)$, 0.31 $(\Box)$, 2.09
$(\diamond)$ and 4.97 $(\times)$. b) Corresponding functions
$S(W_\tau)$. c) The slope $\Sigma(\tau)$ of $S(W_\tau)$ is plotted
versus $\tau$ ($\Box$: experimental values; continuous line:
theoretical prediction from refs. \cite{Douarche2006,Joubaud2006})
with no adjustable parameter).} \label{fig:SSFT}
\end{figure}

\subsection{Transient fluctuation theorem}
We consider first the TFT. The probability density functions
$P(W_\tau)$ of $W_\tau$ are plotted in Fig.\,\ref{fig:TFT}a) for
different values of $\tau$. We see that the PDFs are Gaussian for
all $\tau$ and the mean value of $W_\tau$ is a few $k_B T$. We
also notice that the probability of having negative values of $W$
is rather high for the small $\tau$. The function
\begin{equation}
S(W_\tau) \equiv  \ln\left[{P(W_\tau)\over P(-W_\tau)}\right] \label{eq_FT}
\end{equation}
is plotted in fig.\ref{fig:TFT}b). It is a linear function of
$W_\tau$ for any $\tau$, that is $S(W_\tau)=\Sigma(\tau)~W_\tau$.
Within experimental error, we measure the slope $\Sigma(\tau)=1$.
%
Thus for our harmonic oscillator the TFT is verified for any time
$\tau$. This was expected~\cite{Evans:Searles,Cohen1}, and a
derivation of this generic result for a second order Langevin
dynamics is given in ref.\cite{Douarche2006,Joubaud2006}.

\subsection{Stationary state fluctuation theorem}
We now consider the SSFT with $t_i \ge 3\tau_\alpha$ in
eq.\ref{work}. We find that the PDFs of $W_\tau$, plotted in
Fig.\ref{fig:SSFT}a), are Gaussian with many negative values of
$W_\tau$ for short $\tau$. The function $S(W_\tau)$, plotted in
Fig.\ref{fig:SSFT}b), is
  still a linear function of $W_\tau$, but, in contrast to TFT,  the slope
  $\Sigma(\tau)$ depends  on $\tau$. In Fig.\ref{fig:SSFT}c) the
  measured values of $\Sigma(\tau)$ are plotted as a function of
  $\tau$. The function $\Sigma(\tau)\rightarrow 1$ for
  $\tau \gg \tau_\alpha$. Thus SSFT is verified only for large $\tau$.
The finite time corrections of SSFT, which present oscillations
whose frequency is close to $f_0$, agree quite well with the
  theoretical prediction computed for a second order Langevin
  dynamics \cite{Douarche2006}.
   We stress that the  finite time correction is in
  this case very different from that computed in ref.\cite{Cohen1,Cohen}
  for the first order Langevin equation.
The results of Figs.\ref{fig:TFT},\ref{fig:SSFT} have been checked
for several $M_o/\tau_r$ without noticing any difference. The
errors bars in the figures are within the size of the symbols, and
they come only from the calibration errors of the harmonic
oscillator parameters, and statistics of realizations (typically
$5\times10^5$ cycles have been used).
\begin{figure}
\centerline{\includegraphics[width=0.85\linewidth]{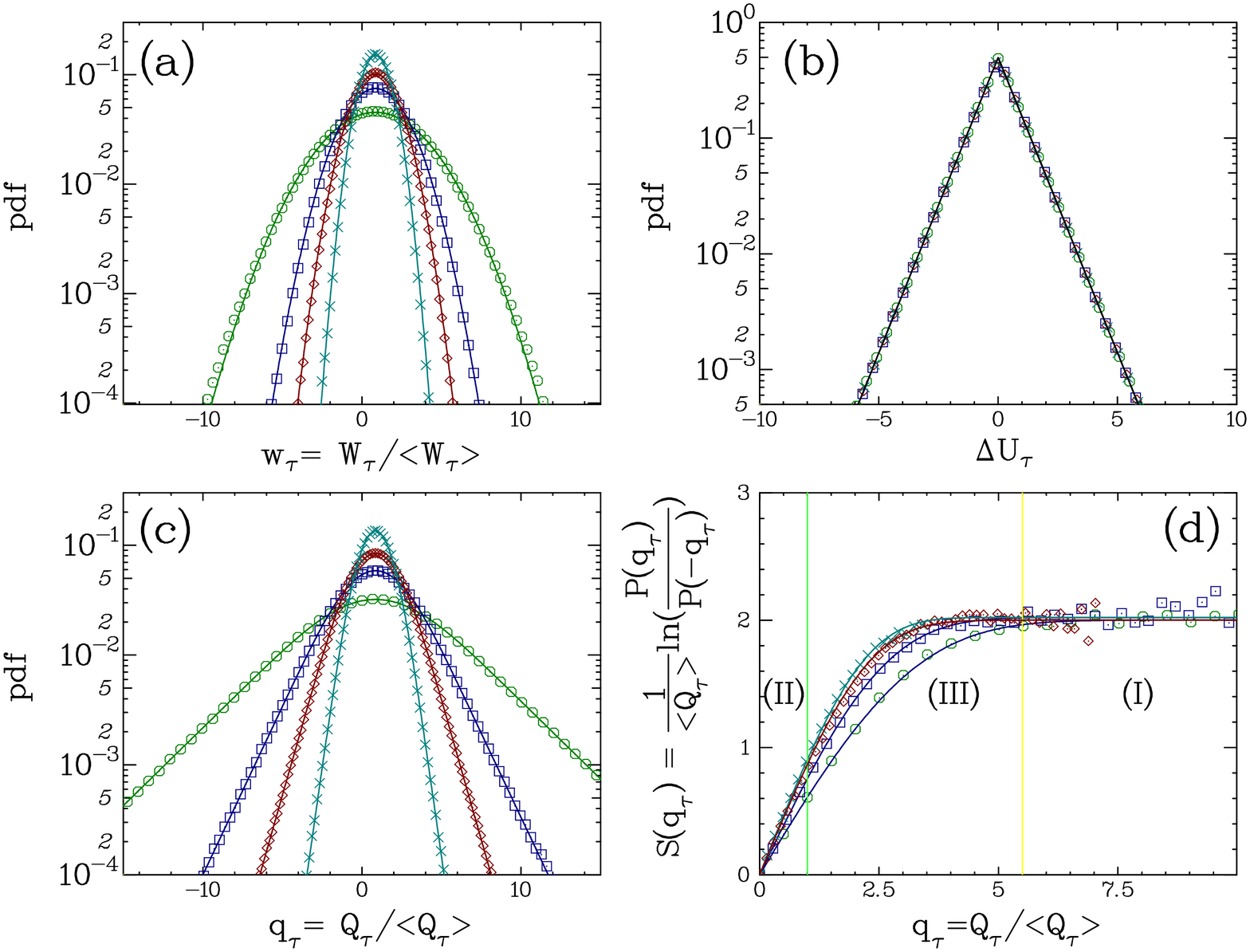}}
\centerline{\includegraphics[width=0.85\linewidth]{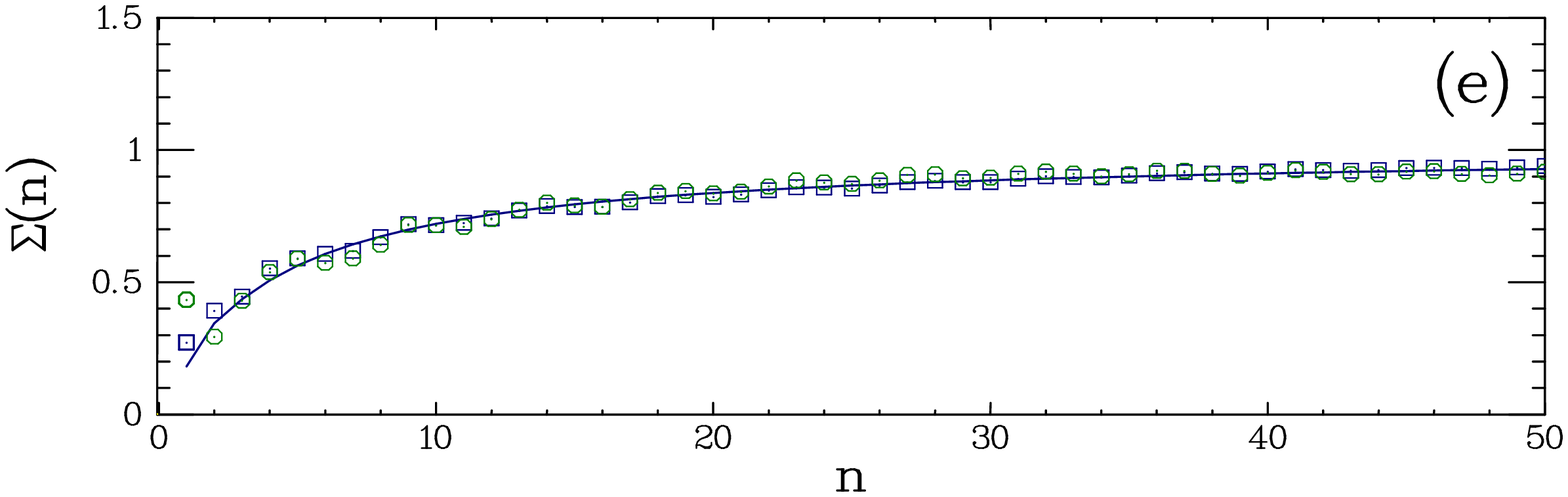}}
\centerline{\includegraphics[width=0.85\linewidth]{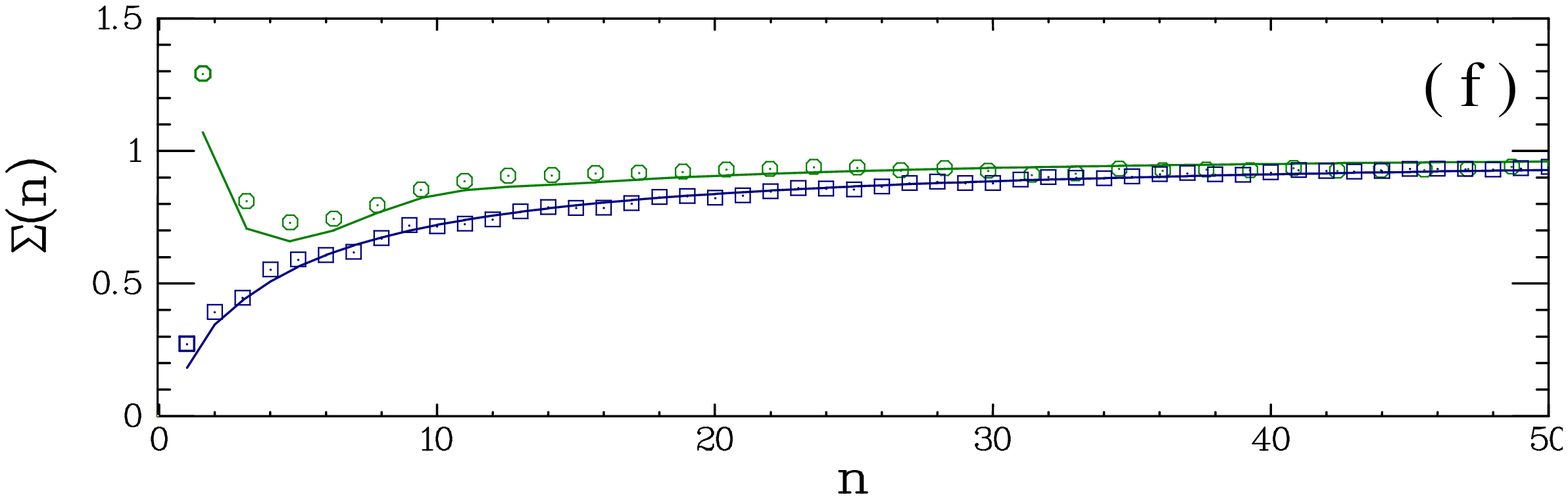}}
\caption{Sinusoidal forcing. a) PDFs of $w_\tau$. b) PDFs of
$\Delta U_\tau$. c) PDFs of $q_\tau$. d) Functions $S(q_\tau)$. In
all plots, the integration time $\tau$ is a multiple of the period
of forcing, $\tau = 2 n \pi/\omega_d$, with $n=7$ ($\circ$),
$n=15$ ($\Box$), $n=25$ ($\diamond$) and $n=50$ ($\times$).
Continuous lines are theoretical predictions with no adjustable
parameter.  e) The slope $\Sigma_q(n)$ of $S(q_\tau)$ for $q_\tau
<1$, plotted as a function of $n$ ($\circ$). The slope
$\Sigma_w(n)$ of $S(w_\tau)$ plotted as a function of $n$($\Box$).
Continuous line is theoretical prediction. f) The slopes
$\Sigma(n)$, plotted as a function of $n$ for two different
driving frequencies $\omega_d$ = 64 Hz ($\Box$) and 256 Hz
($\circ$); continuous lines are theoretical predictions from
ref.\cite{Douarche2006,Joubaud2006} with no adjustable parameter.}
\label{fig:sinusFT}
\end{figure}

\section{Periodic forcing}
We describe in this section the results of the periodic forcing.
In this case $M(t)= M_o \sin \omega_d t$ and the work expression
(eq.{\ref{work}}) is replaced by
\begin{equation}
W_n = W_{\tau=\tau_n} = {1 \over k_B \ T} \ \int_{t_i}^{t_i+\tau}
M(t) {d \theta \over dt} dt \,, \label{eq:work:sinus}
\end{equation}
where $\tau$ is a multiple of the period of the driving ($\tau = 2
n \pi/\omega_d$ with $n$ integer). The starting phase $t_i
\omega_d$ is averaged over all possible $t_i$ to increase
statistics. This periodic forcing is a stationary state that has
never been studied in the details in the context of FT. As already
done for the first order Langevin
equation\cite{Cohen1,Cohen,Seifert}, we want to consider here also
the heat fluctuations, that is the energy dissipated by the
oscillator towards the heat bath. Multiplying
eq.~(\ref{eqoscillator}) by $\frac{\rm d \theta}{\rm dt}$ and then
integrating between $t_i$ and $t_i+\tau$, we obtain exactly the
first law of thermodynamics. The change in the internal energy
$\Delta U_\tau$ over a time $\tau$ is:
\begin{equation}
\Delta U_\tau = U(t_i+\tau) - U(t_i) = Q_\tau + W_\tau
\label{Energyconservation}
\end{equation}
where $Q_\tau$ is the heat given to the system. Equivalently,
$(-Q_\tau)$ is the heat dissipated by the system. For a harmonic
oscillator described by a second order Langevin equation
(eq.\ref{eqoscillator}) the internal energy $U(t)$ and the heat
$Q_\tau$  have the following expressions:
\begin{equation}
U(t)={1 \over k_B \ T} \left[{1 \over 2} I_{\mathrm{eff}}\left[
\frac{\rm d \theta (t)}{\rm dt} \right]^{2} +{1 \over 2} C
\theta(t)^2 \right] \label{Udef}
\end{equation}

\begin{equation}
Q_\tau = \Delta U_\tau - W_\tau = -{1 \over k_B \ T }
\int_{t_i}^{t_i+\tau} \nu \left[ \frac{\rm d \theta}{\rm dt}
(t')\right]^{2}dt' + {1 \over k_B \ T } \int_{t_i}^{t_i+\tau}
\eta(t') \frac{\rm d \theta}{\rm dt}(t') dt' \,. \label{Qdef}
\end{equation}
The first term in eq.\ref{Qdef} corresponds to the viscous
dissipation and is always positive, whereas the second term can be
interpreted as the work of the thermal noise which have a
fluctuating sign.

We rescale the work $W_\tau$ (the heat $Q_\tau$) by the average
work $\langle W_\tau \rangle$ (the average heat $\langle Q_\tau
\rangle$) and define: $w_\tau = \frac{W_\tau}{\langle W_\tau
\rangle}$ ($q_\tau = \frac{Q_\tau}{\langle Q_\tau \rangle}$).
Averages are time-averages, and they are proportional to $\tau$ on
the stationary state under consideration.

As an example we consider a measurement performed at $M_o=0.78pN \
m$ and at $\omega_d/(2\pi)=64Hz$. The PDFs of $w_\tau$, $\Delta
U_\tau$ and $q_\tau$ are plotted in Fig.~\ref{fig:sinusFT} for
different values of $n$. The average of $\Delta U_\tau$ is clearly
zero because the time $\tau$ is a multiple of the period of the
forcing. The PDFs of the work (fig.~~\ref{fig:sinusFT}a) are
Gaussian for any $n$ whereas the PDFs of heat fluctuations
$q_\tau$ have exponential tails(fig.~~\ref{fig:sinusFT}c). These
exponential PDFs can be understood taking into account that, from
eq.~(\ref{Qdef}), $-Q_\tau=W_\tau-\Delta U_\tau$ and that $\Delta
U_\tau$ has an exponential PDF independant of n
(fig.~\ref{fig:sinusFT}b). In the case of the periodic forcing the
PDF of $q_\tau$ can be exactly computed (see
ref.\cite{Joubaud2006}):

\begin{eqnarray} P(Q_\tau) =  {1 \over 4}\exp \left(
\frac{\sigma_W ^2}{2}\right) \left[ e^{Q_\tau - \langle Q_\tau
\rangle} {\rm erfc} \left(\frac{Q_\tau-\langle Q_\tau
\rangle+\sigma_W^2}{\sqrt{2\sigma_W^2}}\right) +  e^{-(Q_\tau -
\langle Q_\tau \rangle)}{\rm erfc} \left(\frac{-Q_\tau+\langle
Q_\tau \rangle+\sigma_W^2} {\sqrt{2\sigma_W^2}}\right)\right] \,,
\label{PDFq}
\end{eqnarray}
where erfc stands for the complementary Erf function and
$\sigma_W^2$ is the work variance. In Fig.~\ref{fig:sinusFT}c, we
have plotted the analytical PDF from eq.~(\ref{PDFq}) together
with the experimental ones, using the values of $\sigma_W^2$ and
$\langle Q_\tau \rangle$ from the experiment. The agreement is
perfect for all values of $n$ and with no adjustable parameter,
using eq.~(\ref{PDFq}).

To quantify the symmetry of the PDF around the origin, we define
the function $S$ as:
\begin{equation}
S(e_\tau) \equiv  {1 \over \langle E_\tau
\rangle}\ln\left[{P(e_\tau)\over P(-e_\tau)}\right]
\label{eq_FT_p}
\end{equation}
where $e_\tau$ stands for either $w_\tau$ or $q_\tau$ and $E_\tau$
stands for either $W_\tau$ or $Q_\tau$. The question we ask is
whether:
\begin{equation}
\lim_{\tau \rightarrow \infty} S(e_\tau) = e_\tau
\label{conventionalFT}
\end{equation}
as required by SSFT. $S(q_\tau)$ is plotted in
Fig.~\ref{fig:sinusFT}d) for different values of $n$ ; three
regions appear:

\begin{enumerate}

\item For large fluctuations $q_\tau$, $S(q_\tau)$ equals $2$.
When $\tau$ tends to infinity, this region spans from $q_\tau = 3$
to infinity.

\item For small fluctuations $q_\tau$, $S(q_\tau)$ is a linear
function of $q_\tau$. We then define $\Sigma_q(n)$ as the slope of
the function $S(q_\tau)$, {\it i.e.} $S(q_\tau) = \Sigma_q(n) \,
q_\tau$. This slope is plotted in Fig.~\ref{fig:sinusFT}e where we
see that it tends to $1$ when $\tau$ is increased. So, SSFT holds
in this region II which spans from $q_\tau=0$ up to $q_\tau = 1$
for large $\tau$.

\item A smooth connection between the two behaviors.

\end{enumerate}

The PDF of the work being Gaussian, the functions $S(w_\tau)$ are
proportional to $w_\tau$ for any $\tau$, {\it i.e.}
$S(w_\tau)=\Sigma_w(n) \, w_\tau$ (ref.~\cite{Douarche2006}).
$\Sigma_w(n)$ is plotted in Fig.~\ref{fig:sinusFT}e) and we
observe that it matches experimentally $\Sigma_q(n)$, for all
values of $n$. So the finite time corrections to the FT for the
heat are the same than the ones of FT for
work~\cite{Douarche2006}: $\Sigma_w(n)=\Sigma_q(n)=1+K/n+1/n\ {O}
\left({\rm e}^{-\tau_n/\tau_\alpha}\right)$, where $K$ is a
constant.

We find that that the  scenario plotted in
fig.\ref{fig:sinusFT}a)-e) is the same for any driving frequency
$\omega_d$
 \cite{Douarche2006,Joubaud2006}.
  However the finite time
 corrections of $\Sigma_w(n)$ are a function of $\omega_d$.
The function $\Sigma_w(n)$, measured at $\omega_d/2\pi=64$Hz,   is
compared with that measured at $\omega_d/2\pi=256$Hz in in
fig.\ref{fig:sinusFT}f). The convergence rate is quite different
in the two cases, which agree with the theoretical predictions for
a second order Langevin equation (see
ref.\cite{Douarche2006,Joubaud2006}).

We also notice that in the case of the sinusoidal forcing, the
convergence is very slow in Figs.~\ref{fig:sinusFT}e) and
~\ref{fig:sinusFT}f), we see that it takes several dozens of
excitation period (500 ms for $\omega_d/2\pi=64$Hz) to get
$\Sigma(n)=1$ within one percent. On the contrary for a ramp
forcing, this is achieved in about (20 ms), i.e. a few
$\tau_\alpha¤$ \ (see Fig.~\ref{fig:SSFT}c).

In this section we have seen that also in the case of the
sinusoidal forcing the agreement between the computed and measured
finite time corrections is very good. These results prove not only
that FTs asymptotically hold for any kind of forcing, but also
that finite time corrections strongly depend on the specific
dynamics. The detailed theoretical analysis of this behavior for a
second order Langevin equation can be found in
ref.\cite{Joubaud2006}.

\begin{figure}[!h]
    \begin{center}
    \hskip 4cm      (a) \hskip 7cm (b) \\
    \includegraphics[width=7cm, angle=0]{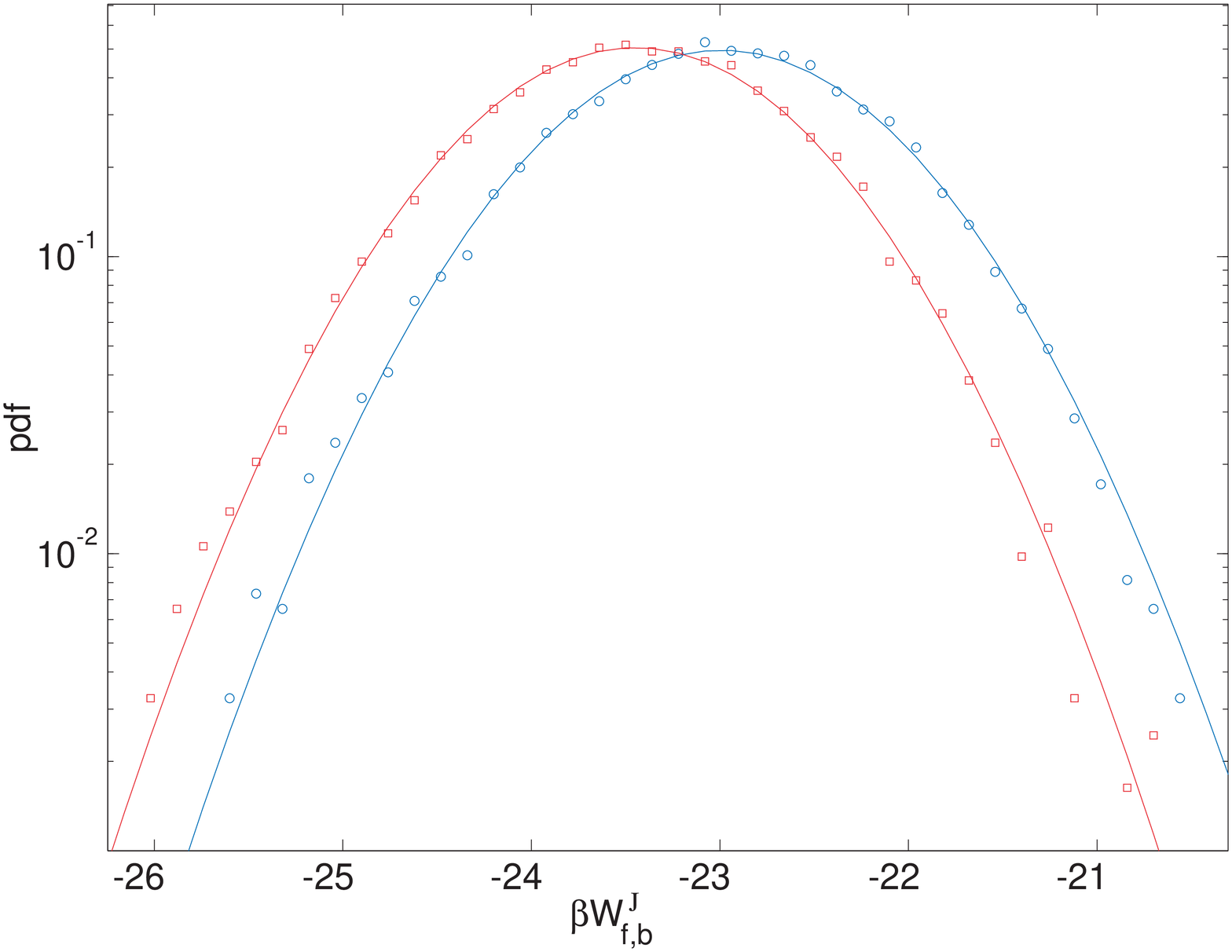}
    \includegraphics[width=7cm, angle=0]{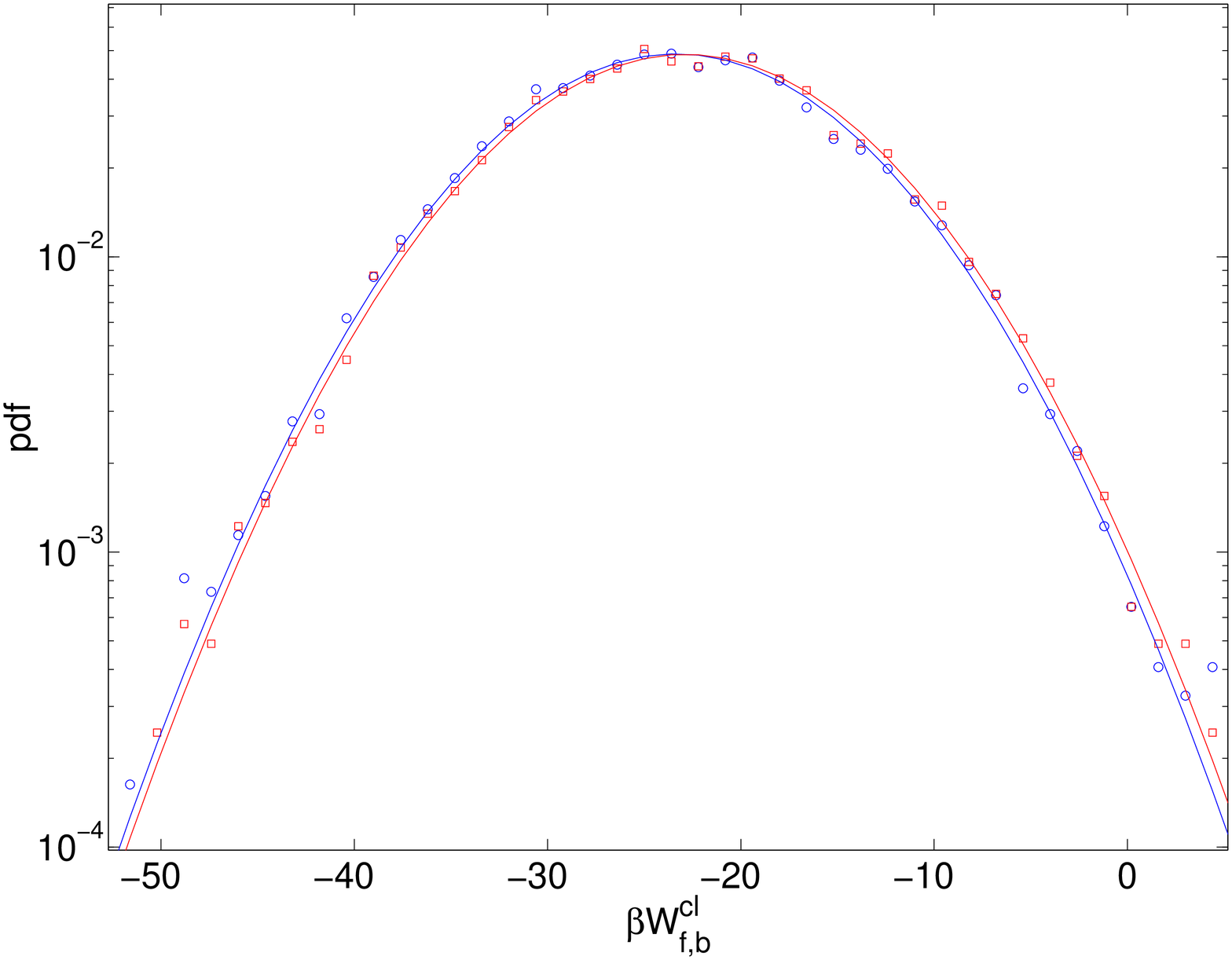}\\
\hskip 4cm      (c) \hskip 7cm (d) \\
    \includegraphics[width=7cm, angle=0]{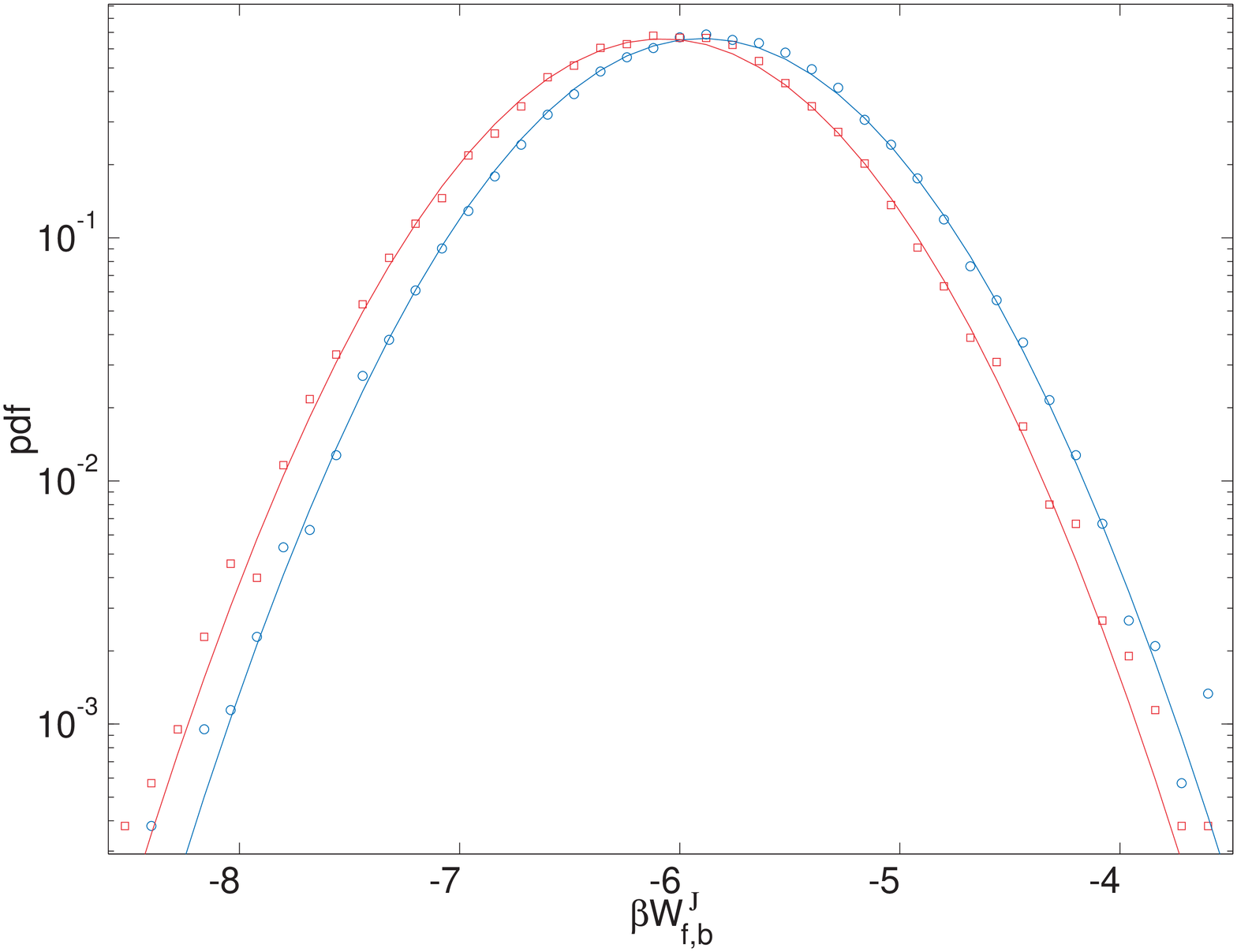}
    \includegraphics[width=7cm, angle=0]{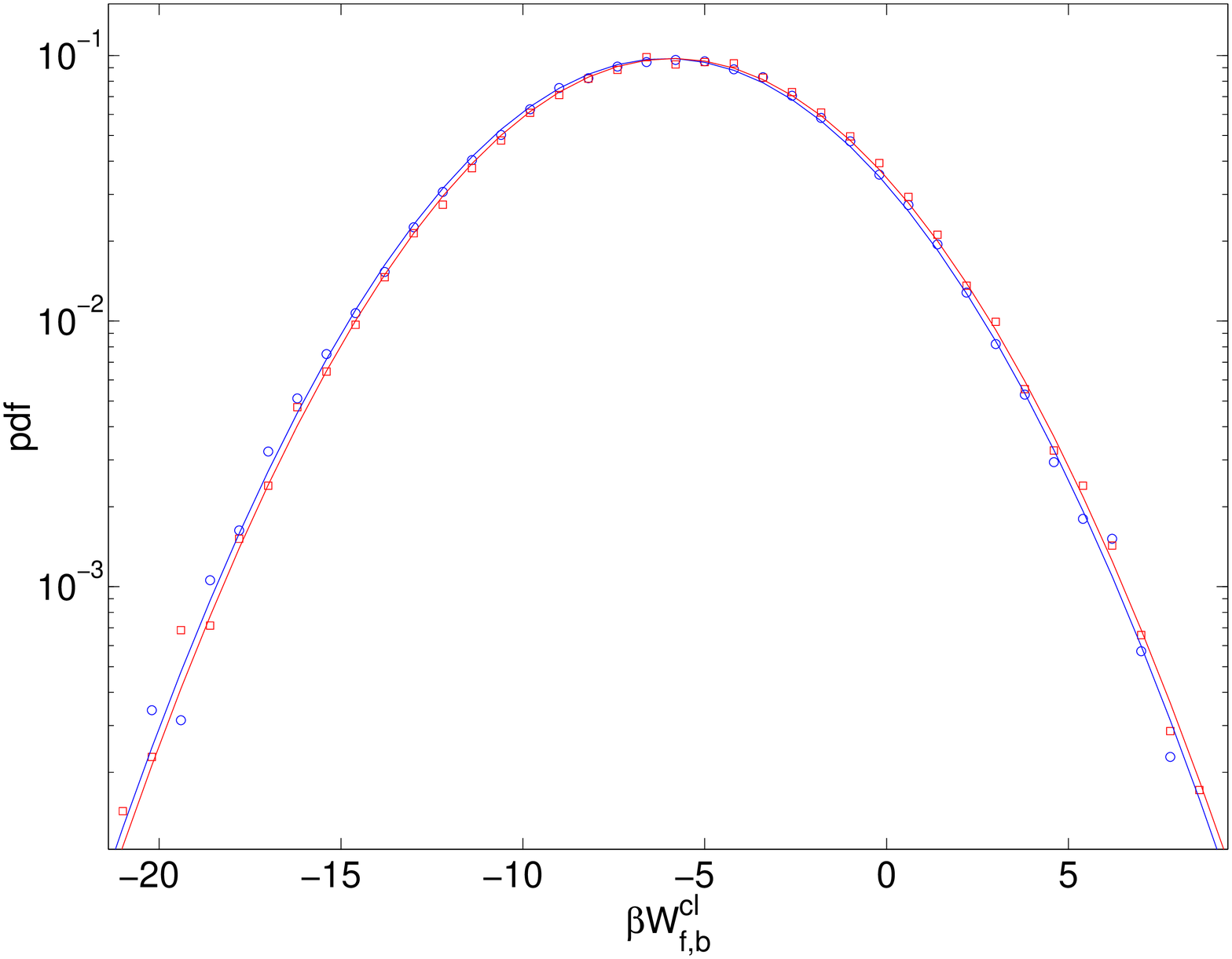}
    \caption{ $M_o=11.9 pN m$ (a) $\mathrm{P}_{\mathrm{f}}(W)$ and $\mathrm{P}_{\mathrm{b}}(-W)$,
    (b) $\mathrm{P}_{\mathrm{f}}(W^{\mathrm{cl}})$ and $\mathrm{P}_{\mathrm{b}}(-W^{\mathrm{cl}})$;
    $M_o=6.1 pN m$: (c) $\mathrm{P}_{\mathrm{f}}(W)$ and $\mathrm{P}_{\mathrm{b}}(-W)$,
    (d) $\mathrm{P}_{\mathrm{f}}(W^{\mathrm{cl}})$ and $\mathrm{P}_{\mathrm{b}}(-W^{\mathrm{cl}})$
    (experimental forward and backward pdfs are represented by $\circ$ and $\Box$ respectively,
    whereas the continuous lines are Gaussian fits)}
    \label{pdfs}
    \end{center}
\end{figure}

\section{TFT and the  Jarzynski equality the Crooks relation}

In this section we want to discuss the relationship between TFT
and JE.  The Jarzynski equality (JE) relates the free energy
difference $\Delta F$ between two equilibrium states $A$ and $B$
of a system in contact with a heat reservoir to the PDF of the
work performed on the system to drive it from $A$ to $B$ along any
path $\gamma$ in the system parameter space. Actually thermal
fluctuations are here very useful because they allow the system to
explore all possible paths from A to B.
Specifically, let us consider the work done by the external torque
on the harmonic oscillator described in  section
\ref{section_ramp}. The external torque $M(t)$ drives the system
from $M(0)=0$ (equilibrium state A) to $M(t>\tau_r)=M_o$
(equilibrium state B). Jarzynski defines
 the work performed by $M(t)$ to drive  the system from A to B as
\begin{equation}
    W^J = -\int_{0}^{\tau_r} \dot{M} \theta \d t =
    -\Bigl[M \theta \Bigr]_{0}^{\tau_r} - W^{\mathrm{cl}},
\label{JEwork}
\end{equation}
where
\begin{equation}
     W^{\mathrm{cl}} = - \int_{0}^{\tau_r} M \dot{\theta} \d t = -
     W_{\tau_r}
     \label{classicalwork}
\end{equation}
is the classical work such that  $- W^{\mathrm{cl}}/(k_BT)$ is
equal to  $ W_{\tau_r}$ defined in eq.\ref{work}. Here we define
$W^{\mathrm{cl}}$ such that it has the same sign of $W^J$. Thus
$W^J$ and $W^{\mathrm{cl}}$ are related but they are not the same
and they differ by a boundary therm, which makes an important
difference in the fluctuations of these two quantities. This
difference between the $W$ and $W_{cl}$, has been already pointed
out in ref.\cite{Hummer}.
 One can consider an ensemble
of realizations of the ''switching process'' from A to B  with
initial conditions all starting in the same initial equilibrium
state. Then the work may be computed for each trajectory in the
ensemble. The JE states that \cite{jarzynski1}

\begin{equation}
    \Delta F = -\frac{1}{\beta} \log \langle \exp{[-\beta W^J]} \rangle,
    \label{JE}
\end{equation}

where $\langle{\cdot}\rangle$ denotes the ensemble average,
$\beta^{-1} = k_B T$. In other words $\langle \exp{[-\beta
W_{\mathrm{diss}}]} \rangle = 1$, since we can always write $W=
\Delta F+ W_{\mathrm{diss}}$ where $W_{\mathrm{diss}}$ is the
dissipated work. In our experiment we can also check the CR which
is related to the JE and which gives useful and complementary
information on the dissipated work. Crooks considers the forward
work $W_{\mathrm{f}}$ to drive the system from $A$ to $B$ and the
backward work $W_{\mathrm{b}}$ to drive it from $B$ to $A$. If the
work pdfs during the forward and backward processes are
$\mathrm{P}_{\mathrm{f}}(W^J)$ and
$\mathrm{P}_{\mathrm{b}}(-W^J)$, one has \cite{crooks1,jarzynski2}
\begin{equation}
    \frac{\mathrm{P}_{\mathrm{f}}(W^J)}{\mathrm{P}_{\mathrm{b}}(-W^J)}
    = \exp{(\beta [W^J-\Delta F])}
    = \exp{[\beta W_{\mathrm{diss}}]}
    \label{crooks}.
\end{equation}
A simple calculation from Eq.\,(\ref{crooks}) leads to
Eq.\,(\ref{JE}). However, from an experimental point of view this
relation is extremely useful because one immediately sees that the
crossing point of the two pdfs, that is the point where
$\mathrm{P}_{\mathrm{f}}(W^J) = \mathrm{P}_{\mathrm{b}}(-W^J)$, is
precisely $\Delta F$. Thus one has another mean to check the
computed free energy by looking at the pdfs crossing point
$W_{\mathrm{\times}}$.

Many details of the test of the JE and CR in the harmonic
oscillator can be found in ref.\cite{DouarcheJSM}. Here we want
only to stress the difference between the fluctuations of the
classical and the Jarzinsky works. This can be seen in
fig.\ref{pdfs} where we plot the the pdfs of $W^J$
(figs.\ref{pdfs} a,c) and $W^{\mathrm{cl}}$ (figs.\ref{pdfs} b,d)
for the forward and backward processes measured for two values of
$M_o$ specifically $M_o=11.9pNm$ (figs.\ref{pdfs}a,b) and
$M_o=6.1pNm$ (figs.\ref{pdfs}a,b). The stiffness C of the
oscillator was $7.5 \ 10^{-4} N m \ rad^{-1}$ in this experiment.
In Figs.\,\ref{pdfs}, bullets are the experimental data and the
continuous lines their fitted Gaussian pdfs. Comparing the pdfs of
$W^J$  and $W^{\mathrm{cl}}$ for the same $M_o$ we see that the
variance of $W^{\mathrm{cl}}$ is much larger than that of $W^J$.
Indeed $W^{\mathrm{cl}}$ presents both positive and negative
values and, as we have seen in sect.\ref{section_ramp},~\
$P(-W^{\mathrm{cl}})$ satisfies TFT. Thus it is clear that if
$W^J$ satisfies eq.\ref{JE} this equation cannot be satisfied by
$W^{\mathrm{cl}}$. Indeed using the measured values of $W^J$ in
eq.\ref{JE} one finds $\beta \Delta F= -23.3\pm 0.4$ for
$M_o=11.9$ and $\beta \Delta F= -6.3\pm 0.3$ for $M_o=6.1$. These
values correspond to $\Delta F = -\frac{M_{o}^2}{2C}$ for the
driven harmonic oscillator. In Fig.\,\ref{pdfs}, the pdfs
$\mathrm{P}_{\mathrm{f}}(W)$ and $\mathrm{P}_{\mathrm{b}}(-W)$
cross at $\beta W \simeq -23.5$ (fig.\ref{pdfs}a) and in $\beta W
\simeq -6.1$ (\ref{pdfs}c), which again correspond to the expected
values. Thus we see that JE and CR can be safely applied on
experimental data  to measure $\Delta F$ if the work $W^J$ defined
in eq.\ref{JEwork}  is used. In ref.\cite{DouarcheJSM} it has been
shown that this  result is true independently of the ratio $\tau_r
/ \tau_\alpha$ and of the maximum amplitude of $M_o$. This has
been checked at the largest $M_{o}$ and the shortest rising time
$\tau_r$ allowed by our apparatus.

The fact that the JE and CR are satisfied only by $W^J$ but not by
$W^{\mathrm{cl}}$ can be easily understood by noticing that  a
variable that satisfies TFT cannot be used in eq.\ref{JE}. Indeed
one may rewrite eq.\ref{JE} as:

\begin{equation}
\exp{-\beta \Delta F}=\ < \exp(-\beta \ \tilde W ) > = \
\int_{-\infty}^{\infty} \exp(-\beta \ \tilde W )P(\tilde W)
d\tilde W \label{JE1}
\end{equation}

where $\tilde W$ is an energy injected into the system on a time
$\tau$. If $P(\tilde W)$ satisfies TFT then from eq.\ref{eq_FT1}
and eq.\ref{eq_TFT} :
$$P(\tilde W)=\exp(-\beta \ \tilde W ) P(-\tilde W), \  \ \
\forall \tau$$. Replacing this identity in eq.\ref{JE1} one finds:
\begin{equation}
\int_{-\infty}^{\infty} \exp(-\beta \ \tilde W ) \ P(\tilde  W)\
dW= \int_{-\infty}^{\infty} \exp(-\beta \ \tilde W )\ P(-\tilde
W)\ \exp(\beta \tilde W)\ d \tilde W = 1 \label{noJE}
\end{equation}
showing that if  $\tilde W$ satisfies TFT then cannot be used to
compute $\Delta F$. Thus one concludes that as $W^{\mathrm{cl}}$
satisfies TFT it cannot satisfy the JE as it shown in
eq.\ref{noJE}. This observation points out the importance of the
choice of the ''good variables'' in order to use the JE to compute
the free energy difference in a more complex system. Several other
limitations for a safe application of JE and CR on real data have
been already discussed in ref.\cite{DouarcheJSM}.

\section{Conclusions}

In conclusion we have applied the FTs and JE to the  work
fluctuations of an oscillator driven out of equilibrium by an
external force. The experimental results agree with  theoretical
predictions both for TFT and SSFT (eqs.\ref{eq_FT1},\ref{eq_TFT}
and \ref{eq_SSFT}). Indeed we observe that in the transient case
when the  system starts in equilibrium (TFT) the FT
(eq.\ref{eq_FT1}) holds for any time, whereas when the system is
in a stationary out of equilibrium regime (SSFT), eq.\ref{eq_FT1}
holds only asymptotically. The SFFT presents a complex convergence
to the asymptotic behavior which strongly depends on the form of
the driving. The exact formula of this convergence can be computed
using several experimental evidences of the statistics of the
fluctuation. The details of derivation of the finite time
correction and of the PDF for the heat will be the subject of a
long paper \cite{Joubaud2006}. We have also discussed the
importance of the choice of the variable in order to safely apply
JE.  The results reported in this paper are useful for many
applications going from biological systems to nanotechnology,
where the harmonic oscillator is the simplest building block.

\vskip 2\baselineskip

{\bf Acknowledgments}
This work has been partially supported by EEC contract DYGLAGEMM
and ANR-05-BLAN-0105-01.

\end{document}